\documentclass[aps,amsmath,twocolumn,amssymb,floatfixng,showpacs, superscriptaddress,footinbib, longbibliography,nofootinbib]{revtex4-1}

\date{\today}
\usepackage{epsfig}
\usepackage{tabularx}
\usepackage{epstopdf}
\usepackage{subfig}
\usepackage{graphicx}
\usepackage{dcolumn}
\usepackage{bm}
\usepackage{amsmath}
\usepackage[colorlinks,linkcolor=blue,hyperindex,CJKbookmarks]{hyperref}
\usepackage{float}
\usepackage{hyperref}
\usepackage{comment}
\usepackage{xcolor}
\usepackage{makecell}
\usepackage{soul}   %to strike out text
\usepackage{todonotes} % for empty figures
\usepackage{textcomp}

\begin{document}

\title{Skyrmion manipulation and logic gate functionality in transition metal multilayers}

\author{Tamali Mukherjee}
\author{V Satya Narayana Murthy$\S$}
\email{satyam@hyderabad.bits-pilani.ac.in}
\affiliation{Department of Physics, BITS Pilani-Hydrabad Campus, Telangana 500078, India}
\author{Banasree Sadhukhan$\S$, $\P$}
\email{banasree.sadhukhan@mahindrauniversity.edu.in}
\affiliation{Department of Physics, École Centrale School of Engineering,  Mahindra University,  Hyderabad,  Telangana 500043,  India}

\def\thefootnote{$\S$}\footnotetext{Joint last authors}
\def\thefootnote{$\P$}\footnotetext{Corresponding author}

\date{\today}

%--------------------------------------------------------
%--------------------------------------------------------
\begin{abstract}

Magnetic skyrmions, due to their topological stability and high mobility, are strong candidates for information carriers in spintronic devices. To advance their practical applications, a detailed understanding of their nucleation and current-driven dynamics is essential. We investigate the formation and manipulation of skyrmions in a square nano structure (200 $\times$ 200 nm$^{2}$, 1 nm thick) of PdFe/Ir(111) multilayers subjected to nano second current pulses with magnitude ranging from (1-5)$\times$10$^{11}$ A/m$^2$.  Using micromagnetic simulations, we demonstrate controlled motion of skyrmion under different types of spin-transfer torque (STT). The calculated skyrmion Hall angle (SkH) for Slonczewski type STT is ${\theta_{SkH}^{SL}} = 89.53^{\circ}$ for  PdFe/Ir(111) multilayers which ensures the edge accululation of skyrmion like a track within the nano structure and we extend this idea further for different shape engineering of skyrmion in 4d tranisition metal multilayers by manipulating the magnitude and direction of current pulses. Next, we investigate the influence of voltage-controlled magnetic anisotropy ranging from (1.4 - 4.2) $\times$ 10$^6$ J/m$^3$ with external magnetic field B$_{ext}$ = 2 T, and (0 - 2.8) $\times$ 10$^6$ J/m$^3$ with B$_{ext}$ = 3 T respectively,  on skyrmion dynamics for designing anisotropy-engineered barriers to guide their trajectories in PdFe/Ir(111) multilayers.  We use further these barriers to implement basic logic operations, including OR and AND gates,  with skyrmions representing binary states. The calculatd skyrmion Hall angle for Zhang-Li type STT in  PdFe/Ir(111) multilayers is ${\theta_{SkH}^{ZL}} = 3.26^{\circ}$. Consequently, the skyrmions propagate predominantly along the direction of the applied current with minimal deflection, a feature that renders them highly suitable for logic operations. 

\end{abstract}

\maketitle

\section{Introduction}
\label{sec_intro}

\par Magnetic skyrmions are swirling spin structures with nontrivial topology that show great potential for future spintronic devices \cite{fillion2022gate, mishra2025magnetic}. Due to their nano scale size and the ability to be efficiently controlled by electric currents, they offer prospects for high-density data storage and rapid computational operations \cite{KHODZHAEV2025101220, rodrigues2025skyrmions}.  Skyrmion nucleation is possible in ferromagnetic / heavy metal (FM / HM) multilayers by applying an external magnetic field, nano second current pulse or ultra-fast laser etc. \cite{dohi2022thin, sadhukhan2024engineering, kandukuri2021isolated, mukherjee2025role, kern2022tailoring, gerlinger2021application, 10.21468/SciPostPhys.18.2.064}. The first evidence of nano scale skyrmions was reported in an Fe monolayer grown on Ir(111) substrate \cite{heinze2011spontaneous} . Since then, thin films with strong interfacial Dzyaloshinskii–Moriya interaction (iDMI) capable of stabilizing high-density,  nano scale skyrmions have been extensively investigated.\cite{fert2013skyrmions, back20202020, kang2016skyrmion}.

\par In the new era of technology, skyrmionic devices offer miniaturized, energy-efficient, non-volatile data storage as a better alternative to the conventional electronic approach. In these devices, skyrmion plays a major role in carrying the information encoded in binary form as `0' (the absence of skyrmion) or `1' (the presence of skyrmion) \cite{tomasello2014strategy, zhu2017skyrmion, zhang2015magnetic, sisodia2022programmable}. As a first step in designing a skyrmionic device, the skyrmions' motion must be well-studied and regulated in a desired fashion to keep the skyrmions at their designated places.  In the thin film multilayer, applying the current to move the skyrmions \cite{iwasaki2013current, iwasaki2013universal, ohki2024fundamental, nie2025current} is feasible in three different ways within micromagnetic simulations of skyrmions : injection of a nano second current pulse a) to the free magnetic layer , b) to the spacer layer providing the required spin-orbit torque (SOT) \cite{woo2017spin, ryu2020current, shao2021roadmap}, or c) to the fixed layer providing the essential spin-transfer torque (STT) \cite{ralph2008spin, komineas2015skyrmion, masell2020spin}.

\par As magnetic skyrmions continue to attract significant interest in spintronics, their controlled manipulation under nano second electrical current pulse is being extensively investigated for prospective device applications. In a typical magnetic tunnel junction configuration, comprising a fixed (reference) ferromagnetic layer and a free magnetic layer separated by an insulating barrier,  the charge current flowing through the fixed layer becomes spin-polarized upon traversing the reference layer. When this spin-polarized current is injected into the free layer, its non-collinear spin component exerts a STT on the local magnetization via exchange interaction. Depending on the strength and duration of the current pulse, this torque can drive either steady-state magnetization precession or full magnetization reversal in the free layer. Specifically, the application of a nano second current pulse in the fixed layer generates a spin-polarized electron flux that transfers angular momentum to the magnetic moments in the free magnetic layer, thereby inducing reorientation of the local magnetization.

\par From the initial development of spintronic devices-such as racetrack memory \cite{tomasello2014strategy, zhu2017skyrmion, kang2019comparative} to more advanced paradigms like neuromorphic computing \cite{li2017magnetic, li2021magnetic, yokouchi2022pattern},  extensive theoretical and experimental efforts have significantly broadened the landscape of spintronic applications. Among various control strategies, modulation of magnetic anisotropy and engineering of anisotropy barriers have emerged as powerful tools for enabling logic functionalities, allowing for reconfigurable and energy-efficient logic operations \cite{luo2018reconfigurable, luo2021skyrmion, wats2024skyrmion}. Building on these foundational approaches, recent advances have turned toward skyrmion-based logic architectures, which exploit the topologically protected nature and current-driven mobility of skyrmions. These nano scale spin textures provide a robust platform for realizing logic operations within magnetic racetracks and junction geometries, further enhancing the potential for dense, low-power spintronic computing.

\begin{figure*}[ht]
\includegraphics[width=0.8\textwidth,angle=0]{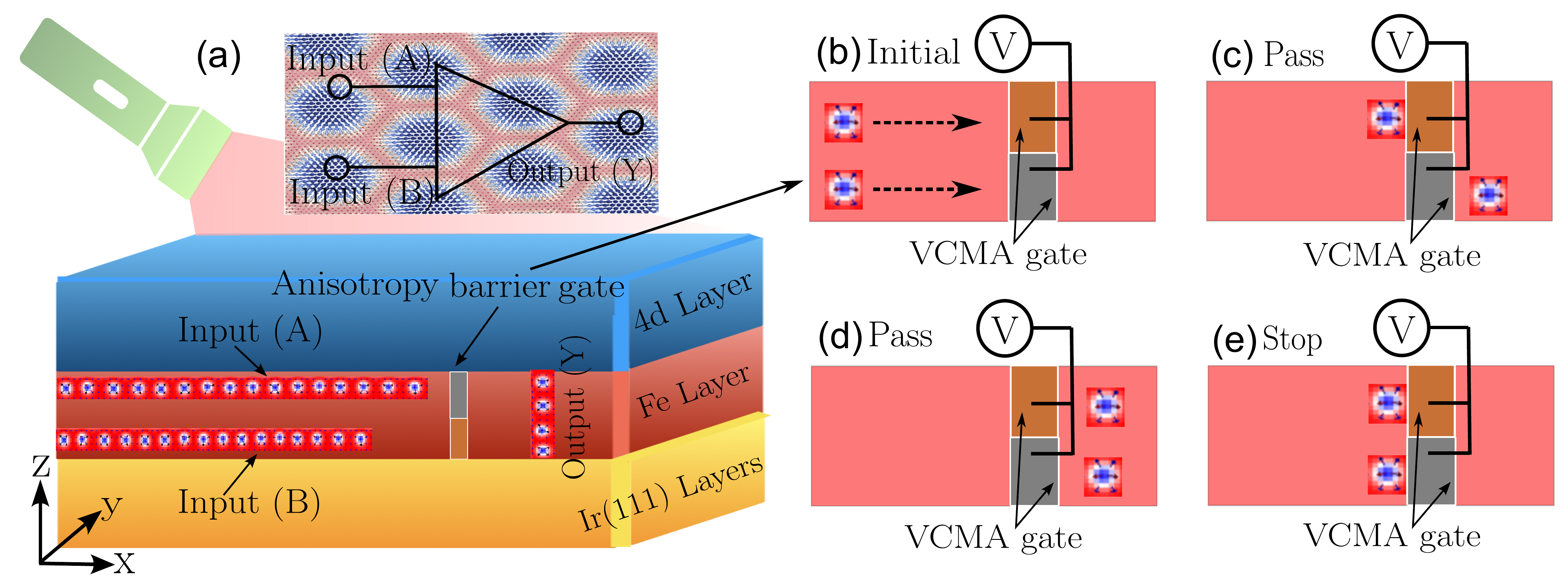}
\caption{ (a)-(e) Schematic representation of voltage-controlled magnetic anisotropy (VCMA) based skyrmion logic gate in 4d transition metal multilayers. The pass and stop conditions of skyrmion in logic gate depends on VCMA gate voltage.}
\label{fig1}
\end{figure*}

\par Magnetic racetracks, junctions, and engineered nano structures are commonly utilized to direct the motion and facilitate controlled interactions of skyrmions, forming the foundational elements of skyrmion-based logic gates \cite{luo2018reconfigurable, Paikaray_2023}. In a typical realization of an AND gate, two skyrmions introduced simultaneously at distinct input branches interact such that a single skyrmion emerges at the output, thereby fulfilling the logical AND condition. Conversely, an OR gate can be implemented by designing the geometry such that the presence of a skyrmion at either input path leads to its detection at the output. These functionalities are often achieved through collision and merging dynamics within Y-shaped junctions \cite{Paikaray_2023} or by guiding skyrmions along converging nano tracks \cite{xia2018skyrmion} . The NOT gate operation is realized by ensuring a complementary response where the presence of a skyrmion at the input corresponds to its absence at the output, and vice versa. This can be accomplished via skyrmion-skyrmion repulsion or by exploiting branched geometries that redirect or annihilate the skyrmion depending on the input condition. Such gate architectures demonstrate the feasibility of constructing fundamental logic operations using skyrmionic degrees of freedom.

\par In this work, we propose a novel design for skyrmion-based logic gates that harness voltage-controlled magnetic anisotropy (VCMA) to manipulate skyrmion trajectories without the need for high current densities in a  PdFe/Ir(111) system.  By engineering anisotropy landscapes within magnetic multilayers, we demonstrate how electric fields can dynamically reconfigure the energy barriers that guide skyrmion motion and perform logic operations such as OR and AND.  Micromagnetic simulations reveal that VCMA-driven gates can operate with high speed and energy efficiency while maintaining non-volatility.  Our approach offers a scalable and programmable route toward integrating skyrmionics into low-power, high-density logic circuits for future computing architectures. The manuscript is organized as follows: Sec.\ref{sec_method} describes the computational method followed; Sec.\ref{sec_result1} and \ref{sec_result2} first focuses on the nucleation, manipulation of skyrmions in detail and then how this controlled motion of skyrmion lead to logic gate functionality with VCMA.  Finally,  Sec.\ref{sec_conclu} summarizes the notable outcomes and concludes with insights into potential directions for further study.

\section{Methodology and Micromagnetic Simulations}
\label{sec_method}

\par The nucleation and subsequent manipulation of skyrmions in magnetic multilayers can be realized through two distinct mechanisms: (i) an in-plane spin-polarized current (CIP) injected into the ferromagnetic layer, or (ii) an electrical current driven through HM layer, which via the spin Hall effect generates a spin current polarized perpendicular to the plane (CPP).  For simulations of the STT associated with the CIP and CPP scenarios, the Zhang and Li (${\tau_{STT}}={\tau_{ZL}}$),  and Slonczewski (${\tau_{STT}}={\tau_{SL}}$), STT terms, respectively, were added to the  Landau-Lifshitz-Gilbert (LLG) equation.  Both kinds of STT are implemented in the simulation package \textsc{Mumax}$^3$~\cite{vansteenkiste2014design, leliaert2019tomorrow}.  The LLG equation along with both STT and SOT is given by :
\begin{equation}
\frac{d \bm{m}}{d t}=
-\gamma \bm{m}\times\bm{H}_{eff}
-\alpha\bm{m} \times \frac{d \bm{m}}{d t}+\bm{\tau}_{STT} + \bm{\tau}_{SOT}
\end{equation}
Where, $\gamma$ = the gyromagnetic ratio and $\alpha$ = Gilbert damping parameter.
The effective magnetic field (H$_{eff}$) has contributions from the external
magnetic field (H$_{ext}$), demagnetizing field (H$_{demag}$), anisotropy field
(H$_{anis}$), and exchange field (H$_{exch}$).   In the present study, we investigate both mechanisms. For skyrmion nucleation, we use only CPP methods.  Whereas the shape engineering of skyrmion has been investigated using both CPP and CIP methods. Finally,  we use CIP method for logic gate formulation of skyrmion in transition metal multilayers.

\par For simulations of STT associated with the CIP scenario, the Zhang and Li type STT term \cite{menezes2019deflection} is given by :
\begin{equation}
\tau_{ZL}= \frac{P\mu_B}{eM_s(1+\beta^2)(1+\alpha^2)}[\bm{m} \times (\bm{m}
 \times (\bm{j}.\nabla)\bm{m})+(\beta - \alpha)\bm{m}\times (\bm{j}.\nabla)\bm{m}] 
 \end{equation}
 Here, P = polarization of the current density applied, $\mu_B$ = Bohr magneton, e = charge of an electron, M$_s$ = satauraion magnetization, $\beta$ = non adiabatic factor, and j = current density applied.  This scenario is directly implemented within \textsc{Mumax}$^3$~\cite{vansteenkiste2014design, leliaert2019tomorrow}.  The Thiele equation for the CIP scenario\cite{zhang2016magnetic} reads : 
\begin{equation}
\bm{G} \times (\bm{v}^{(s)} - \bm{v}^{(d)})+  \mathcal{D} (\beta\bm{v}^{(s)} - \alpha\bm{v}^{(d)}) + \nabla V = 0
\end{equation}
The first term refers to the Magnus force where G= 4$\pi$Q and, the topological charge(Q) is given by, 
\begin{equation}
Q = \frac{1}{4\pi} \int dxdy   \bm{m} . (\partial_x\bm{m}\times \partial_y\bm{m})
\end{equation}
 $\textbf{v}^{(s)}$ = velocity induced by the spin current, $\textbf{v}^{(d)}$ = drift velocity, and $\mathcal{D}$= dissipative tensor, where 
 \begin{equation}
 \mathcal{D}_{ij}= \int dxdy (\partial_i\bm{m}.\partial_j\bm{m})
 \end{equation}
 and $\nabla V$ represents the local potential energy faced by the skyrmion.  The velocity components along x and y direction are as follows \cite{menezes2019deflection, zhang2016magnetic},
\begin{equation}
{v}_{x}^{(d)} = \frac{\mathcal{D}G(\alpha-\beta)}{G^2+\alpha^2\mathcal{D}^2} 
\end{equation}
\begin{equation}
{v}_{y}^{(d)} = \frac{G^2+\mathcal{D}^2\alpha\beta}{G^2+\alpha^2\mathcal{D}^2} 
\end{equation}
As a consequence, the skyrmion Hall angle (${\theta_{SkH}^{ZL}}$) is found to be,
\begin{equation}
{\theta_{SkH}^{ZL}} = tan^{-1}\frac{\mathcal{D}G(\alpha-\beta)}{G^2+\mathcal{D}^2\alpha\beta}
\end{equation}

In the CPP configuration, an electrical current driven through HM layer gives rise to a spin current that is injected into the ferromagnetic layer along the $z$ direction. Within \textsc{Mumax}$^3$~\cite{vansteenkiste2014design, leliaert2019tomorrow, joos2023tutorial},  this situation can be modelled by introducing a fixed layer with polarization vector $\bm{p}$ placed on top of the film, while the applied current is injected along the $z$ direction.  For simulations of STT associated with the CPP scenario, the Slonczewski type STT \cite{joos2023tutorial} term is given by :
\begin{equation}
    \tau_{SL}=\beta' \frac{\epsilon-\alpha \epsilon'}{1+\alpha^2}(\bm{m}\times(\bm{p}\times\bm{m}))-\beta' \frac{\epsilon'-\alpha\epsilon}{1+\alpha^2}(\bm{m}\times\bm{p})
\end{equation}
where,
\begin{equation}
    \beta'=\frac{\hbar j_z}{M_{sat}ed}
\end{equation}
and, 
\begin{equation}
    \epsilon = \frac{P\Lambda^2}{(\Lambda^2+1)+(\Lambda^2-1)(\bm{m}.\bm{p})}
\end{equation}

\par Here, P, $\bm{p}$ and j$_z$ corresponds to the spin polarized current, $\Lambda$, d $\epsilon$ refer to properties of the spacer layer and interface, $\hbar$ = reduced Planck's constant. The Thiele equation for the CPP scenario \cite{jiang2017direct, menezes2019deflection, zhang2016magnetic} reads : 
\begin{equation}
-\bm{G} \times  \bm{v}^{(d)} - \alpha  \mathcal{D} \bm{v}^{(d)} + 4\pi \mathcal{B} j_{HM} +\nabla V = 0
\end{equation}
The velocity components along x and y direction are as follows \cite{menezes2019deflection, zhang2016magnetic} :
\begin{equation}
{v}_{x}^{(d)} = \frac{\alpha \mathcal{D}}{G^2+\alpha^2\mathcal{D}^2} 4\pi \mathcal{B}j_{HM}
\end{equation}
\begin{equation}
{v}_{y}^{(d)} = \frac{G}{G^2+\alpha^2\mathcal{D}^2} 4\pi \mathcal{B}j_{HM} 
\end{equation}
Hence, the skyrmion Hall angle reads,
\begin{equation}
{\theta_{SkH}^{SL}} = tan^{-1}\frac{G}{\alpha \mathcal{D}}
\end{equation}

\par We consider a square nano structure of PdFe/Ir(111) with dimension of (200 $\times$ 200) nm$^{2}$ to study the nucleation and dynamics of skyrmions under the application of nano second current pulse.  The LLG equation is solved through it by employing the finite difference discretization of the sample into (256 $\times$ 256 $\times$ 1) cells.  No thermal fluctuations are taken into account in the course of the simulation time.  The material parameters used to carry out the simulation are M$_{s}$ = 6.3 $\times$ 10$^{5}$ A/m, $\alpha$ = 0.023, exchange constant (A$_{ex}$) = 2.269 $\times$ 10$^{-12}$ J/m, interfacial DMI (D$_{int}$) = 3.64 $\times$ 10$^{-3}$ A/m$^{2}$, first order anisotropy constant (Ku$_{1}$) = 1.4 $\times$ 10$^{6}$ J/m$^{3}$ and easy axis is taken along (0, 0, 1) direction, and free layer thickness is taken as 1 nm.

\par Slonczewsi-type STT parameters used are,  $\Lambda$ = 1.0, P = 0.5 and $\epsilon'$ = 0. The charge current here is in \textbf{j$_z$} = (0, 0, 1) direction, injected through fixed layer having magnetization in \textbf{p} = (1, 0, 0) direction.  We further examine the application of SOT in the sample within the framework of Slonczewski torque \cite{joos2023tutorial},  the significant parameters we use, are spin Hall angle ($\alpha_{H}$) = 0.5, the thickness of the spacer layer (d) = 1 nm, and the direction of the magnetic moments injected are \textbf{p} = (0, 1, 0) as the injected charge current is \textbf{j$_{HM}$} = (1, 0, 0), injected through the heavy metal layer.  For Zhang Li type STT, the current is injected in the free layer, in $\pm$ x and $\pm$ y direction where $\beta$ = 0.2 is considered.  For any practical future applications, the effect of geometry on skyrmion nucleation and dynamics must be analyzed. In this context, we take up a circular disc of 200 nm diameter and a rectangular track of area (200 x 20) nm$^2$ and thickness = 1 nm and analyze the result obtained for the same.

\section{Skyrmion nucleation with magnetic field and current pulse} 
\label{sec_result1}

\subsection{With magnetic field} 

\par PdFe/Ir(111) transforms into a stable skyrmion state from the spin spiral phase in the presence of an external magnetic field applied perpendicularly to the plane of the system \cite{romming2013writing}.  We observed that a short duration of the dc field ($<$ 2.5 ns) is enough for the nucleation of the skyrmion state.  First, the spin spiral ground state gradually transforms into a mixed state having skyrmions in the spin-spiral background then to skyrmions in the ferromagnetic background and finally to a complete ferromagnetic state with the increasing strength of the magnetic field.  Magnetic field ranging between 1.7 T to 4.5 T can form and hold the skyrmions in the ferromagnetic background.  With the increasing the magnitude of the applied magnetic field,  the nucleation threshold time and the resulting skyrmions' radii decrease while the skyrmion count increases, as explained in our previous work \cite{mukherjee2025interplay}.

\subsection{With current pulse} 

\par
\begin{figure}[ht]
\includegraphics[width=0.5\textwidth,angle=0]{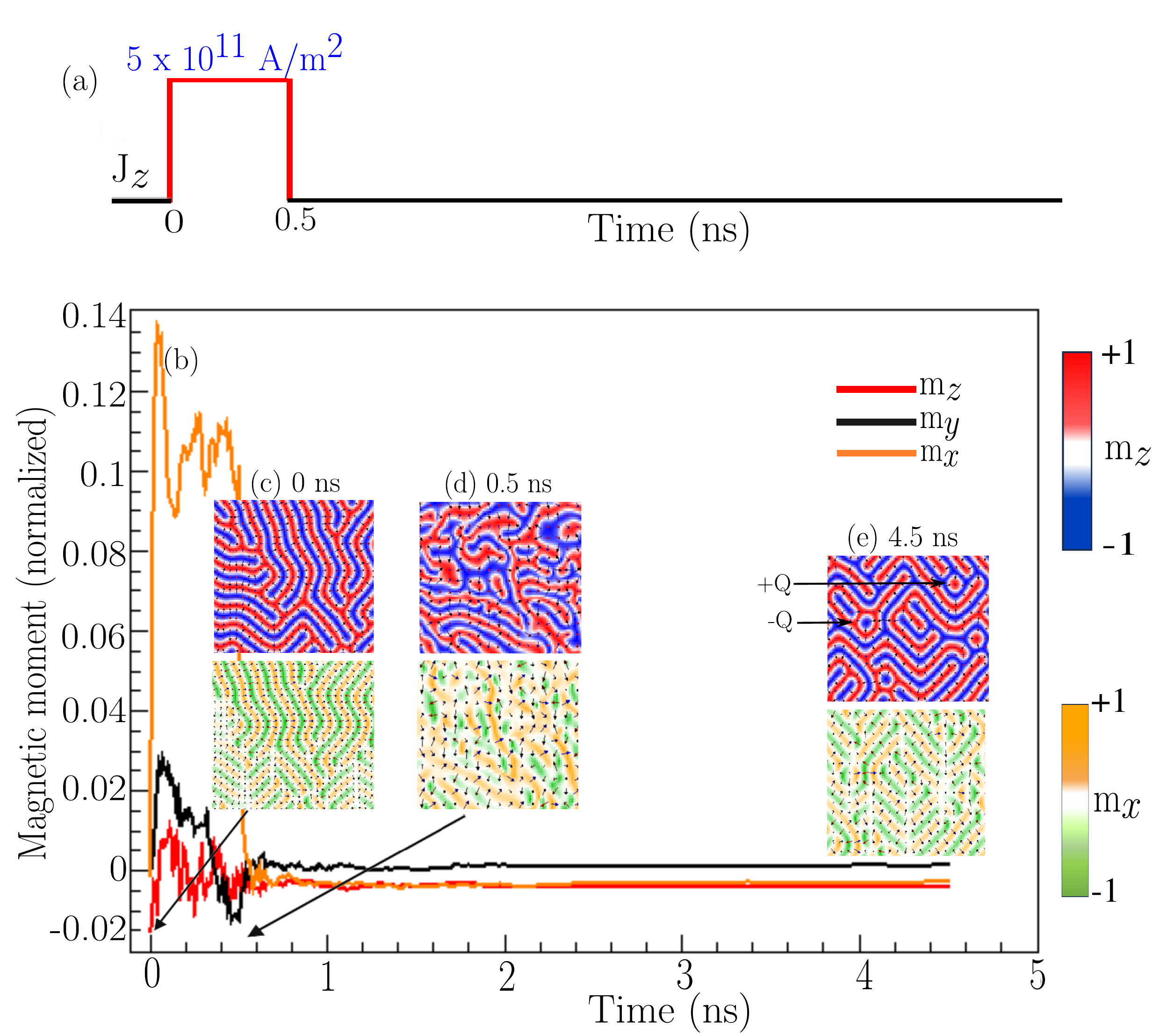}
\caption{ (a) Current pulse of magnitude 5 x 10$^{11}$ A/m$^2$ in +z -- direction for 0.5 ns.  (b) Variation of x, y, and z components of the total magnetic moment with the simulation time.  Zoomed snapshots of the magnetic moments of the nano structure with respect to m$_x$ and m$_z$ components are taken at (c) initial state (d) 0.5 ns, (e) 4.5 ns.}
\label{fig2}
\end{figure}

 As an alternative approach other than the well-studied method of applying an external magnetic field to form skyrmions in PdFe/Ir(111), we used a nanosecond current pulse through the fixed ferromagnetic layer having magnetization in (1, 0, 0) direction. Starting from a current pulse of the order of 10$^{11}$ A/m$^2$, the initial state of the spin-spiral begins its transformation into a mixed phase of skyrmions of opposite core magnetization (skyrmion number $Q=\pm 1$) in a spin-spiral background.  \ref{fig2}(a) shows the current pulse of 5 x 10$^{11}$ A/m$^2$ in +z -- direction for 0.5 ns applied in PdFe/Ir(111).   \ref{fig2}(b) illustrate the variation of in-plane and out of plane components of magnetic moment of the system as time proceeds.  The STT experienced by the Fe spins affects the in-plane component of the magnetic moment (m$_x$) in transient state.  The spin-spiral ground state of the system finally relaxes to a mixed state of skyrmions in the spin-spiral background,  as depicted in \ref{fig2}(e).  

\par The final relaxed state of the system is a stable skyrmion phase with opposite polarity due to the out of plane component magnetization ($m_z$) which nucleates skyrmion in PdFe/Ir(111) under the application of current pulse.  Reversing the direction of the applied current pulse does not yield any new outcome; the system still stabilizes into skyrmions of opposite topological charge (Q = $\pm 1$) within a spin-spiral background.  With the increase in the strength of the current pulse,  even at current pulse strengths approaching 10$^{16}$ A/m$^2$,  the system fails to fully transition into a homogeneous skyrmion phase.

%%%%%%%%%%%%%%%%%%%%%%%%%%%%%%%%%%%%%%%%%%%%%%%%%%%%%%%

\section{Current driven dynamics of skyrmion with magnetic field} 
\label{sec_result2}

\subsection{Edge accumulation of skyrmion}

\begin{figure*}[ht]
\includegraphics[width=0.99\textwidth,angle=0]{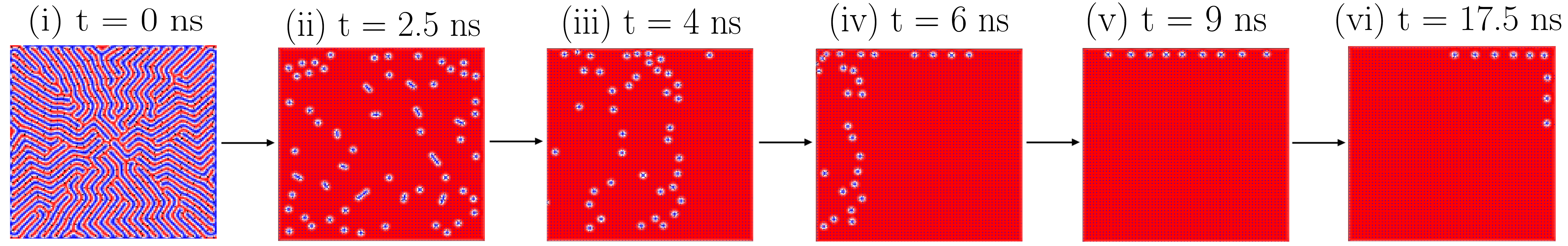}
\caption{A current pulse of 5 $\times$ 10$^{11}$ A/m$^2$ is injected through fixed layer for 10 ns. The snapshots are (i) spin-spiral ground state, (ii) stable skyrmion phase obtained after applying $\vec{B}_{\mathrm{ext}}$ = (0, 0, 2) T for 2.5 ns. (iii)-(v) explains the dynamics of the skyrmions that survive when the current pulse is on, and (vi) the relaxed motion of the skyrmions along the edge when current pulse is off. }
\label{fig3}
\end{figure*}

\par To manipulate skyrmion dynamics under nano second current pulses, we perform micromagnetic simulations in presence of external magnetic field.  Initially,  a out-of-plane magnetic field of 2~T is applied along the $+z$ direction for 2.5~ns, which drives the transformation of the spin-spiral ground state into a skyrmion texture embedded in a ferromagnetic background. To induce current-driven dynamics, a spin-polarized current pulse with a density of $5 \times 10^{11}~\mathrm{A/m^2}$ is applied through the fixed magnetic layer along the $+z$ direction for 10~ns, followed by an additional 5~ns simulation to capture the relaxation behavior after the pulse is switched off.

\par Figure  \ref{fig3} illustrates the temporal evolution of the magnetic state under the influence of both the external field and nanosecond current pulse.  During the current pulse, the skyrmions experience a transverse drift toward the $-x$ direction, driven by the combined action of STT and the emergent Magnus force.  This motion results in their accumulation along the top edge of the nano structure, where they are confined due to edge repulsion and the absence of an energy barrier.  After the pulse is removed, the skyrmions at the boundary exhibit a slow, relaxed clockwise motion along the edge.  These simulations underscore the importance of pulse timing, edge effects, and intrinsic skyrmion topology in controlling their motion for potential device applications.

\par The observed edge accumulation, with nano second current pulse in presence of external magnetic field, is attributed due to the Slonczewski type STT generated in the PdFe/Ir(111) magnetic multilayer.  Upon reversing the direction of the external magnetic field,  a current density of the same magnitude is applied along the $+z$ direction which drives the skyrmions toward the $+x$ direction, resulting in their accumulation at the lower edge of the sample.  In the absence of any current pulse, the relaxed motion of the skyrmions is observed along the edge in an anti-clockwise trajectory.  Owing to their non-zero topological charge,  skyrmions experience a Magnus force, which causes them to move at a finite angle with respect to the applied current direction due to the skyrmion Hall angle, $\theta_{\mathrm{SkH}}$.  For a Slonczewski-type STT~\cite{Paikaray_2023, jiang2017direct}, the calculated Hall angle is given by ${\theta_{SkH}^{SL}} = 89.53^{\circ}$ (for skyrmions having diameter $\approx$ 4.5 nm as measured in our case\cite{simon2014formation, romming2013writing}) which validate the top (bottom) edge accumulation of skyrmion while moving along $-x$ ($x$) directions in presence of J$_{z}$ = $5 \times 10^{11}~\mathrm{A/m^2}$.  By exploiting these edge-residing skyrmions, we demonstrate the formation of distinct geometrical patterns through controlled tuning of the current density direction and magnitude.

\par  We further perform an numerical experiment by applying nano second current pulse to Ir(111) layer to investigate whether SOT can accumulate skyrmion at the edge of PdFe/Ir(111).  Unlike STT,  SOT enables magnetization manipulation without requiring current flow through the ferromagnetic layer,  thus offering improved energy efficiency and endurance.  Additionally,  compared to STT driven motion,  SOT driven skyrmion dynamics are more robust, enabling higher driving currents and correspondingly higher velocities, particularly in racetrack geometries.  Therefore,  SOT has emerged as an efficient mechanism for driving magnetic skyrmions.  This makes SOT particularly attractive for applications in next-generation magnetic memory and logic devices, where deterministic switching can be achieved by engineering structural asymmetry or employing external symmetry-breaking fields.   SOT ($\tau_{SOT}$) has been implemented as an an additional torque in \textsc{Mumax}$^3$ and has the same mathematical form as $\tau_{SL}$.  With the Slonczewski type STT parameters mapped through the derived relations, the existing STT framework can be directly used to simulate SOT in PdFe/Ir(111) multilayers.

\par Strong SOT in PdFe/Ir(111) arises from the interplay between spin-orbit coupling and broken inversion symmetry in multilayered systems.  The interfaces for Pd-Fe bilayer on Ir(111) is responsible for the large DMI,  which stabilizes skyrmions, simultaneously give rise to a strong spin-orbit coupling.  When an in-plane charge current is applied within the Ir(111) layer,  spin-orbit interactions generate a transverse spin current via spin Hall effect.  The resulting spin accumulation at the PdFe/Ir(111) interface diffuses into the ferromagnetic layer,  exerting a torque on its local magnetization.  Generated SOT can be decomposed into two symmetry-allowed components : a damping-like and a field-like torque,  which resembles an effective magnetic field.  Once the complete skyrmion state is obtained after 2.5 ns,  an in-plane current pulse of magnitude 5 $\times$ 10$^{11}$ A/m$^2$ is applied through the Ir(111) layer from the $+x$  direction,  then it provides the equivalent SOT to make the skyrmions move in the system and accumulate at the edge.  Figure \ref{sfig1} in appendix shows the manipulation of skyrmion dynamics under both field and damping like SOT.  This ensures that the field-like SOT is responsible for edge accumulation in PdFe/Ir(111).

%%%%%%%%%%%%%%%%%%%%%%%%%%%%%%%%%%%%%%%%%%%%%%%%%%%%%%%

\subsection{Shape engineering with skyrmion}

\par Manipulating or arranging the skyrmions in a desired pattern is essential for the application to data storage, logic gate, and spintronics. In this section, we describe how skyrmions, initially distributed randomly within the structure, can be arranged in desired patterns by carefully controlling the magnitude and direction of the current pulse. Within a small physical space of the sample, the proper arrangement of the small-sized skyrmions by nano second current pulse opens a new avenue in the application of skyrmions.  A huge number of skyrmions are obtained after application of magnetic field of 3 T for 1 ns and taken to rearrange them in a desired fashion.  If the magnetic field of 3 T is applied in the $z$ direction to the square nano structure,  a total of 286 isolated skyrmions of core magnetization in $z$ direction are nucleated in the ferromagnetic background.  Next we will focus on how to create different shapes by applying the current pulses through the fixed layer and free layer accordingly.

\begin{figure}[ht]
\includegraphics[width=0.5\textwidth,angle=0]{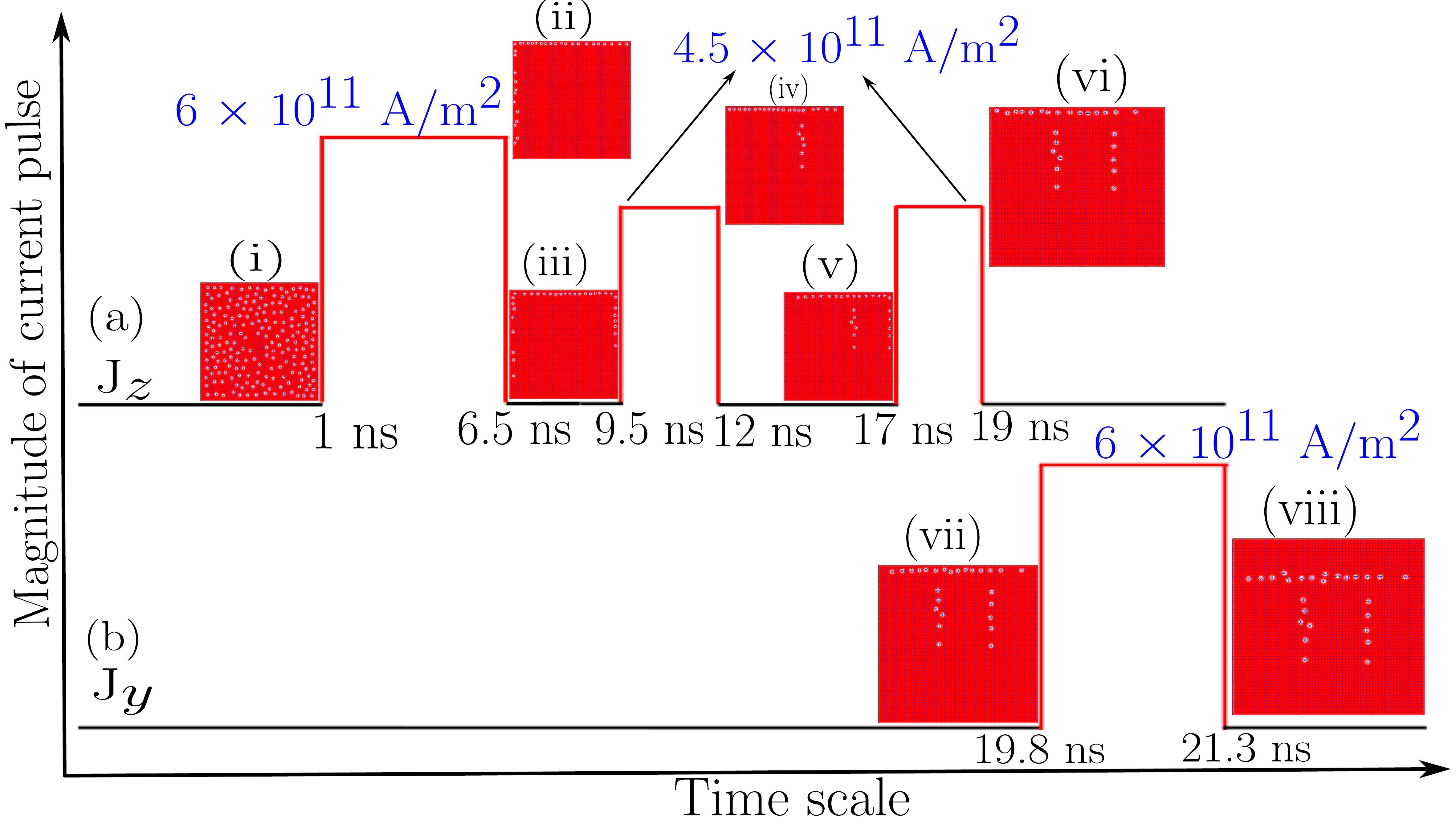}
\caption{Manipulating the skyrmions to form \textbf{`$\pi$'} shape.  (a) and (b) are the current pulses applied along $+z$ and $-y$ directions where (i)- (viii) snapshots during the formation of the \textbf{`$\pi$'} shape. }
\label{fig4}
\end{figure}

\par The `$\pi$'-shaped configuration is generated by a controlled sequence of nano second current pulses applied along the $+z$ and $-y$ directions through the fixed layer and free layer respectively.  As shown in  \ref{fig4},  an initial $6\times10^{11}\,\mathrm{A/m^2}$ pulse in the $+z$ direction for $5.5$~ns,  followed by a $3$~ns interval without current, leaves only the edge skyrmions, which migrate along the sample boundaries.  A subsequent $4.5\times10^{11}\,\mathrm{A/m^2}$ pulse in the $+z$ direction for $2.5$~ns, followed by $5$~ns off,  forms the first vertical segment of the `$\pi$'. The second vertical segment is produced by a $4.5\times10^{11}\,\mathrm{A/m^2}$ pulse for $2$~ns in the $+z$ direction.  Finally, a $6\times10^{11}\,\mathrm{A/m^2}$ pulse in the $+y$ direction for $1.5$~ns centers the entire `$\pi$' structure within the square nano structure.  As demonstrated in the appendix,  appropriate tuning of pulse amplitude, duration, and direction, enables additional geometries such as `T' and `L' shapes in square nano structures,  and `C' and `7' shapes in circular nano structures respectively.

%%%%%%%%%%%%%%%%%%%%%%%%%%%%%%%%%%%%%%%%%%%%%%%%%%%%%%%

\subsection{Departing  from racetrack to logic gate}

\begin{figure}[ht]
\includegraphics[width=0.5\textwidth,angle=0]{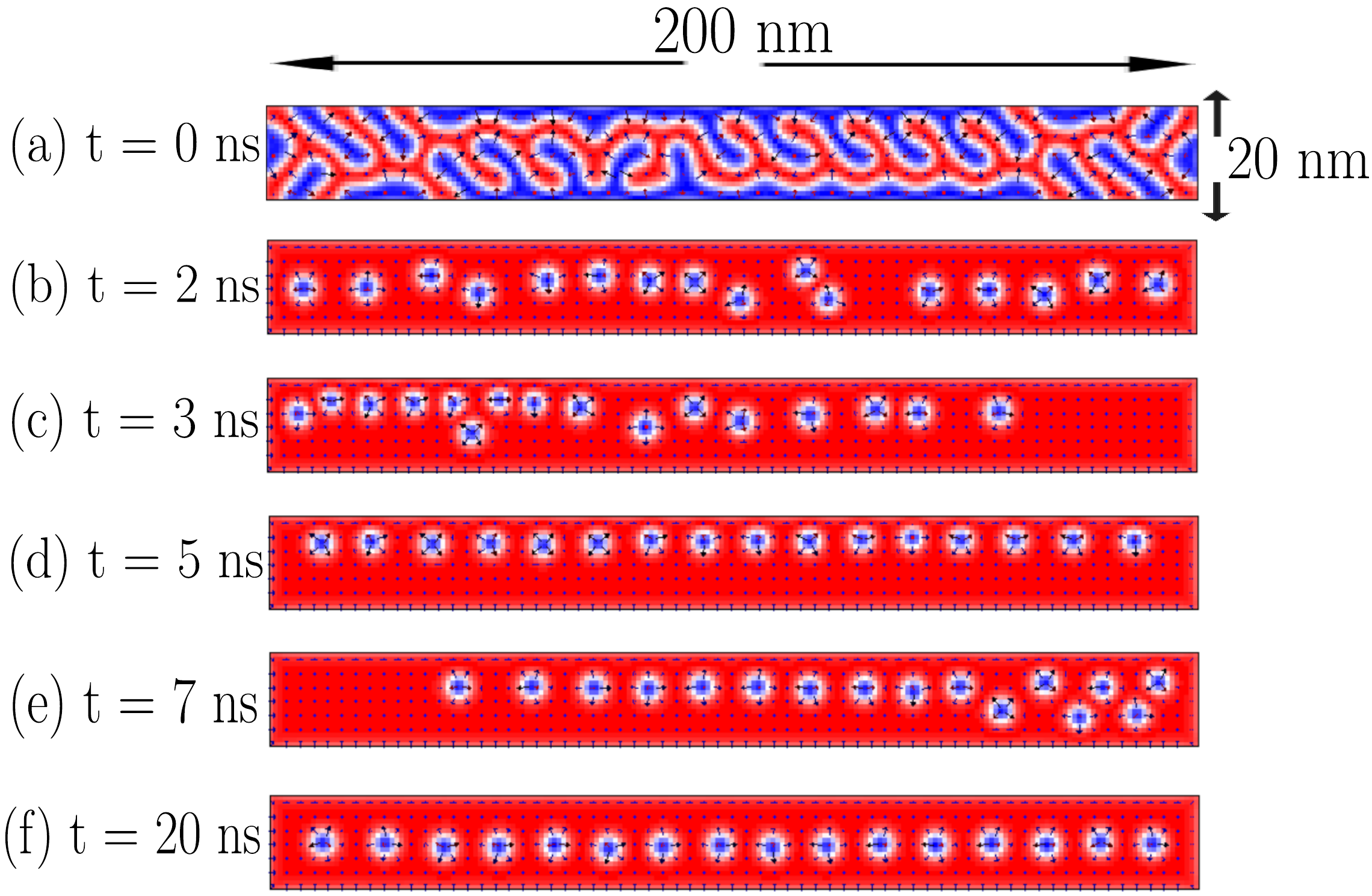}
\caption{A stable skyrmion phase is obtained in (200 $\times$ 20) nm$^2$ rectangular track from (a) spin-spiral ground state. (b) At t = 2ns, 16 isolated skyrmions are nucleated. A current density pulse of 1 $\times$ 10$^{12}$ A/m$^2$ is applied in +x - direction  for 4.5 ns. (c) and (d) are snapshots taken when the current pulse is on. In absence of any current density pulse, from (e) to (f), the skyrmions align in the track and form a racetrack structure. }
\label{fig5}
\end{figure}

\par Since skyrmions can be nucleated in PdFe/Ir(111) multilayers irrespective of the device geometry, their formation is also observed in a rectangular track of dimensions $200\times20\times1~\mathrm{nm^3}$ under an external magnetic field of $2~\mathrm{T}$.  As shown in Fig.\ref{fig5}(a)-(b), the spin-spiral ground state at $t = 0~\mathrm{ns}$ evolves into a skyrmion phase embedded in a ferromagnetic background by $t = 2~\mathrm{ns}$.  Subsequently,  a current pulse of $1\times10^{12}~\mathrm{A/m^2}$ is applied along the $x$ direction for $4~\mathrm{ns}$, driving the skyrmions toward the left edge, where they are repelled and accumulate near the top boundary (see Fig. \ref{fig5}(c)-(d)).  Finally,  it exhibits that after the current pulse is removed,  the skyrmions undergo relaxed motion and eventually settle along the entire length of the track as shown in Fig. \ref{fig5}(e)-(f).

\begin{figure*}[ht]
\includegraphics[width=1.02\textwidth,angle=0]{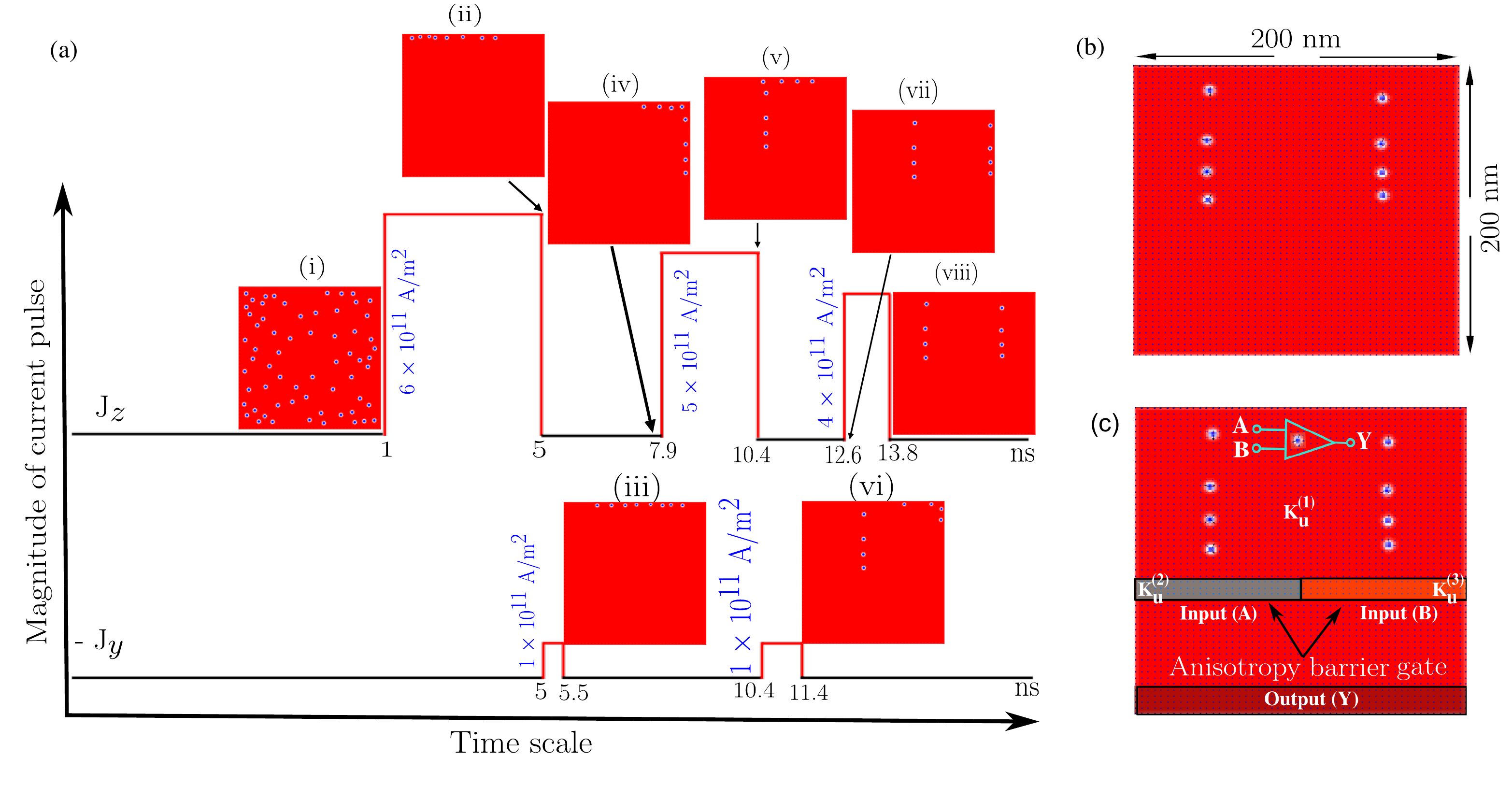}
\caption{(a) Current pulses applied along +z and -y directions to generate the (b) two racetrack-like structures of 4 skyrmions in each line for making (c) logic gate functionality in the square nano structure of dimension (200 $\times$ 200) nm$^2$.}
\label{fig6}
\end{figure*}

\par This concept enables the formation of two vertical skyrmion arrays within a square nanostructure,  serving as inputs for logic gate operations.  In this geometry,  skyrmions with a topological charge of $Q = -1$ are nucleated within $3~\mathrm{ns}$ under an external magnetic field $\mathbf{B}_{\mathrm{ext}} = (0,0,2)~\mathrm{T}$.  Subsequently, current pulses of varying magnitudes and durations are applied along the $+z$ and $-y$ directions, as illustrated in Fig. \ref{fig6}(a).  This pulse sequence generates two racetrack-like skyrmion arrangements which are then utilized to implement OR and AND logic gate functionalities whose schematic representation is shown in Fig. \ref{fig6}(b)-(c).

%%%%%%%%%%%%%%%%%%%%%%%%%%%%%%%%%%%%%%%%%%%%%%%%%%%%%%%

\begin{figure*}[ht]
\includegraphics[width=1.02\textwidth,angle=0]{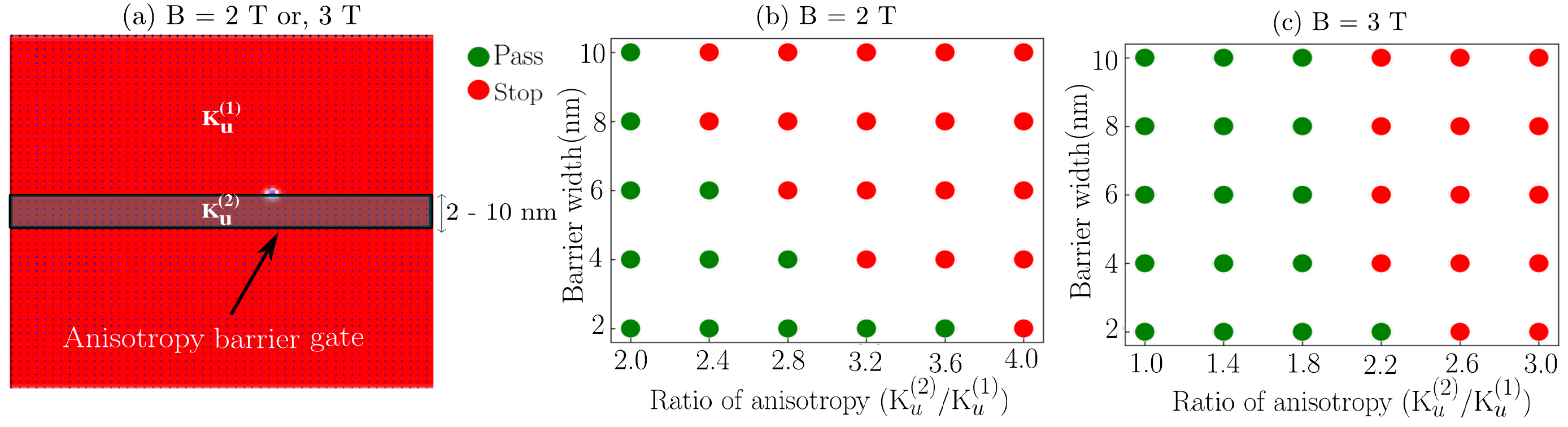}
\caption{ (a) Anisotropy barrier of length 200 nm and width ranging from 2 nm to 10 nm in the square nano structure.  Diagram of skyrmion transmission in terms of anisotropy ratio $K_u^{(2)}/K_u^{(1)}$ and barrier width with external magnetic field (b) $\mathbf{B}_{\mathrm{ext}} = (0,0,2)~\mathrm{T}$ and (c) $\mathbf{B}_{\mathrm{ext}} = (0,0,3)~\mathrm{T}$, respectively.  In all cases, a current pulse of $1\times 10^{12}~\mathrm{A/m^2}$ is applied along $+y$, driving the skyrmion downward along $-y$.
}
\label{fig7}
\end{figure*}

\subsection{Skyrmion filtering via anisotropy barrier}

\begin{figure*}[ht]
\includegraphics[width=0.99\textwidth,angle=0]{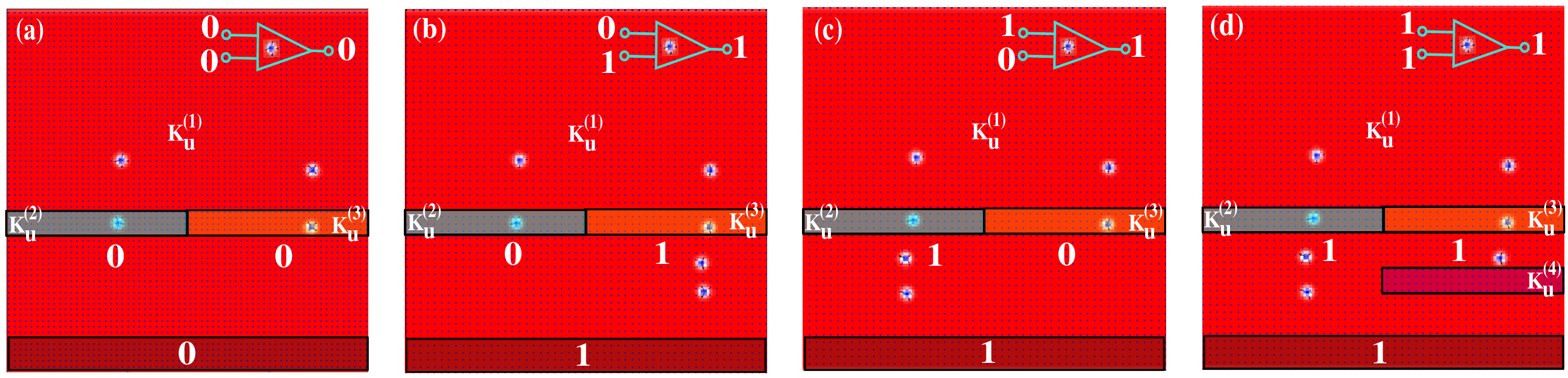}
\caption{ Execution of logic `OR': A Current pulse of (0, 1 $\times$ 10$^{12}$, 0) A/m$^2$ used to move the skyrmions. Three anisotropy barriers used used have anisotropy constants, K$_{u}^{(2)}$, K$_{u}^{(3)}$, and K$_{u}^{(4)}$.  Snapshots are taken for (a) low output, and (b) - (d) high output}
\label{fig8}
\end{figure*}

\par Logic gates are designed by manipulating the dynamics of skyrmions through the application of current pulses in conjunction with voltage-controlled magnetic anisotropy (VCMA).  Within the square nano structure, an anisotropy barrier is employed to regulate and confine the motion of skyrmions, thereby enabling the realization of OR and AND gate functionalities.  For the implementation of these logic gates, the barrier region is engineered to possess an anisotropy constant (K$_{u}^{(2)}$) distinct from that anisotropy of the surrounding nano structure (K$_{u}^{(1)}$ which is anisotropy constant of PdFe/Ir(111)).  K$_{u}^{(2)}$ can be expressed as,  K$_{u}^{(2)}$ = K$_{u}^{(1)}$ + K$_{u}^{(VCMA)}$, where K$_{u}^{(VCMA)}$ we vary in the range of (1.4 - 4.2) $\times$ 10$^6$ J/m$^3$ for B$_{ext}$ = 2 T and (0 - 2.8) $\times$ 10$^6$ J/m$^3$ for B$_{ext}$ = 3 T respectively.

\begin{figure*}[ht]
\centering
\includegraphics[width=0.99\textwidth,angle=0]{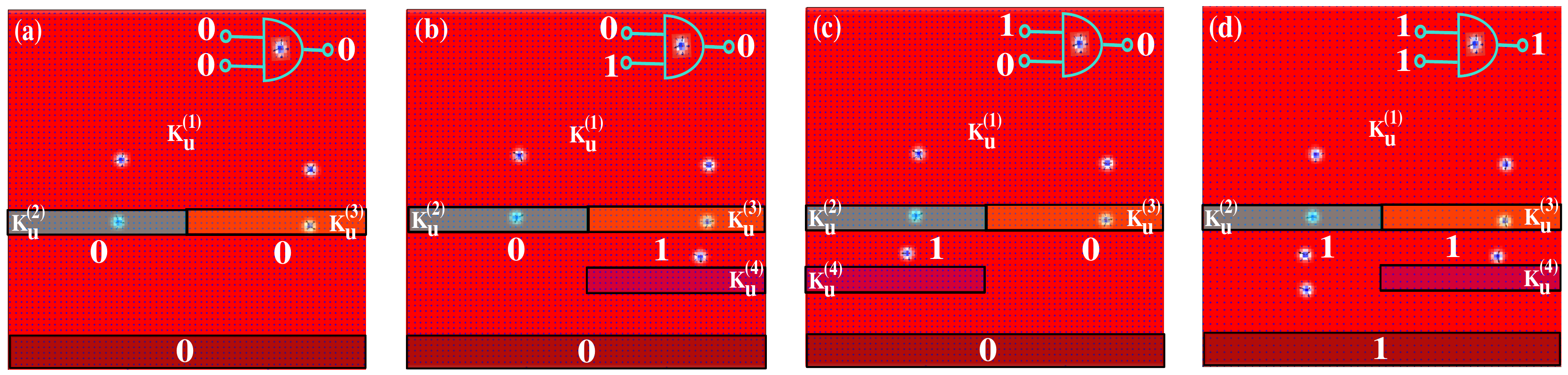}
\caption{ Execution of logic `AND': Current pulse of (0, 1 $\times$ 10$^{12}$, 0) A/m$^2$ used to move the skyrmions. Three anisotropy barriers used have anisotropy constants, K$_{u}^{(2)}$, K$_{u}^{(3)}$, and K$_{u}^{(4)}$.Snapshots are taken for (a) - (c) low output, and (d) high output.}
\label{fig9}
\end{figure*}

\par When a skyrmion encounters the anisotropy barrier,  its ability to pass through or be halted is governed by the barrier strength, which depends on both the anisotropy constant and the barrier width.  We establish the conditions for skyrmion transmission by systematically varying the anisotropy ratio $K_u^{(2)}/K_u^{(1)}$ and the barrier width (in nm).  Figure \ref{fig7}(a)  illustrates a skyrmion approaching a rectangular barrier of fixed length $200~\mathrm{nm}$, with the width varied from $2~\mathrm{nm}$ to $10~\mathrm{nm}$ and positioned at the center of the square nano structure.  Figures \ref{fig7}(b)-(c) present the dependence of the anisotropy ratio $K_u^{(2)}/K_u^{(1)}$ on the barrier width for external magnetic fields $\mathbf{B}_{\mathrm{ext}} = (0,0,2)~\mathrm{T}$ and $\mathbf{B}_{\mathrm{ext}} = (0,0,3)~\mathrm{T}$, respectively.  In both cases, a current pulse of magnitude $1 \times 10^{12}~\mathrm{A/m^2}$ is applied in the $+y$ direction, driving the skyrmion motion vertically downward along the $-y$ direction.

\subsection{Logic gate execution with skyrmion as bit}

\par Binary information is encoded by the presence (‘1’) or absence (‘0’) of a skyrmion.  Two vertical chains of four skyrmions each serve as inputs, with their motion controlled by two anisotropy barriers (length $100~\mathrm{nm}$, width $6~\mathrm{nm}$) positioned at $(50,-30)~\mathrm{nm}$ and $(-50,-30)~\mathrm{nm}$ relative to the  structure center.  For $\mathbf{B}_{\mathrm{ext}} = (0,0,2)~\mathrm{T}$, the barrier anisotropies $K_u^{(2)}$ and $K_u^{(3)}$ are tuned between $2K_u^{(1)}$ and $2.5K_u^{(1)}$ to permit or block skyrmion passage, enabling logic operations.  A current pulse of $1\times 10^{12}~\mathrm{A/m^2}$ along $+y$ drives the skyrmions downward.

%\subsubsection{OR gate}

\par An OR gate performs the logical addition of two binary inputs, $A$ and $B$, yielding the output $Y = A + B$.   Table \ref{tab1} summarizes the required conditions and resulting states for its implementation.   Figure \ref{fig8} illustrates the configuration in which two chains of skyrmions approach the respective anisotropy barriers.  In Fig.\ref{fig8}(a), both anisotropy constants, $K_u^{(2)}$ and $K_u^{(3)}$, are set to $2.5K_u^{(1)}$, preventing skyrmion transmission from either input, thereby realizing $0 + 0 = 0$.  In Fig.\ref{fig8}(b), $K_u^{(3)}$ is reduced to $2.0K_u^{(1)}$, allowing skyrmions from input $B$ to pass while blocking those from $A$, corresponding to $0 + 1 = 1$.  Similarly, Fig.\ref{fig8}(c) shows the case where $K_u^{(2)} = 2.0K_u^{(1)}$ and $K_u^{(3)} = 2.5K_u^{(1)}$, permitting passage from input $A$ only, thus giving $1 + 0 = 1$.

\par Finally,  Fig.\ref{fig8}(d) corresponds to $K_u^{(2)} = K_u^{(3)} = 2.0K_u^{(1)}$, enabling skyrmions from both inputs to cross their respective barriers.  To ensure correct binary output, an additional anisotropy barrier (length $100~\mathrm{nm}$, width $6~\mathrm{nm}$), identical in dimensions to the input barriers, is introduced downstream at either $(50~\mathrm{nm}, -50~\mathrm{nm})$ or $(-50~\mathrm{nm}, -50~\mathrm{nm})$.  Its anisotropy constant is set to $K_u^{(4)} = 2.5K_u^{(1)}$, exceeding both $K_u^{(2)}$ and $K_u^{(3)}$, thereby blocking one skyrmion chain and producing the correct output $1 + 1 = 1$.

\par AND gate is a fundamental logic element that produces a high output ($1$) only when all inputs are high.  
It performs the logical multiplication of two input signals, $A$ and $B$, such that $Y = A \cdot B$.  Table  \ref{tab2} summarizes the input-output conditions and the corresponding observations for the AND gate operation.  Figure \ref{fig9} shows the two input chains of skyrmions, $A$ and $B$.  In Fig.\ref{fig9}(a), two skyrmions from each input chain have already been annihilated at the barriers, while the remaining skyrmions approach barriers of equal strength, $K_u^{(2)} = 2.5K_u^{(1)}$ and $K_u^{(3)} = 2.5K_u^{(1)}$.  Under these conditions, all skyrmions are ultimately annihilated at the barriers, yielding the operation $0 \cdot 0 = 0$.  In the next configuration, $K_u^{(2)}$ is maintained at $2.5K_u^{(1)}$, while $K_u^{(3)}$ is reduced to $2.0K_u^{(1)}$, corresponding to the input state $0$ and $1$.

\par For correct AND gate operation, it must be ensured that no skyrmions remain that could produce a high output when the logical condition is not satisfied.  To this end, an additional anisotropy barrier of dimensions $(100 \times 6)~\mathrm{nm}^2$ and anisotropy constant $K_u^{(4)} = 2.5K_u^{(1)}$ is introduced at $(50~\mathrm{nm}, -50~\mathrm{nm})$ within the square nano structure.  This barrier prevents skyrmions from input $B$ from surviving, resulting in an output of $0$, as shown in Fig. \ref{fig9}(b).   When the input configuration is reversed by setting $K_u^{(2)} = 2.0K_u^{(1)}$ and $K_u^{(3)} = 2.5K_u^{(1)}$, the third barrier with $K_u^{(4)} = 2.5K_u^{(1)}$ is instead placed at $(-50~\mathrm{nm}, -50~\mathrm{nm})$ to block skyrmions from input $A$, again yielding an output of $0$ [Fig.\ref{fig9}(c)].  
Finally, Fig.\ref{fig9}(d) considers the case where both inputs are high, with $K_u^{(2)} = K_u^{(3)} = 2.0K_u^{(1)}$.  Similar to the OR gate scenario, placing the third barrier at either $(50~\mathrm{nm}, -50~\mathrm{nm})$ or $(-50~\mathrm{nm}, -50~\mathrm{nm})$ ensures that only one skyrmion chain-either $A$ or $B$-persists, producing the correct AND gate output of $1$.

%\subsection{Understanding the dynamics of skyrmion}

\begin{figure*}
\includegraphics[width=0.99\textwidth,angle=0]{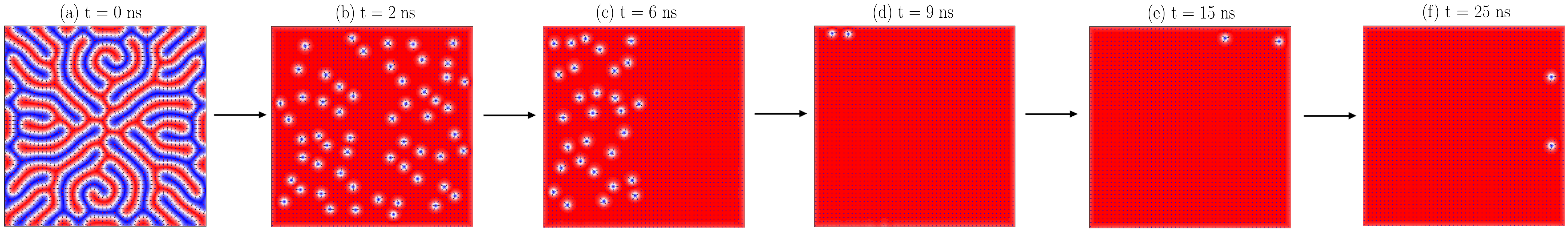}
\caption{Application of current pulse in RuFe/Ir(111).  From (a) the spin-spiral ground state (b) a complete skyrmion phase is obtained after the application of the dc magnetic field of strength of 1.5 T for 2 ns. Then, a nano second current density pulse of 3 $\times$ 10 $^{11}$  A/m$^2$ is applied to the system in +z direction through the fixed layer for 7 ns and switched off. Meanwhile, snapshots are taken at (c) at 6 ns when the current pulse is on, (d) at 9 ns when the current is just switched off; two skyrmions are accumulated at the top edge, (e) at 15 ns when the skyrmions are showing a relaxed motion along the top edge, and (f) at 25 ns the skyrmions' damped relaxed motion along the edge in the clockwise direction. This result resembles the scenario observed in PdFe/Ir(111).} 
\label{fig10}
\end{figure*}

\par Table \ref{tab3} refers to the crossing time for all the aforementioned anisotropy barriers corresponding to the anisotropy constants, K$_{u}^{(2)}$, K$_{u}^{(3)}$ and K$_{u}^{(4)}$ that explore the practical feasibility of a high-speed logic device.  Here, the spin polarized current is in the FM free layer (CIP) and the skyrmions are experiencing Zhang - Li torque. Though the current is applied in the y direction, the skyrmions will have a velocity along x and y direction due to Hall effect.  The calculated Hall angle in CIP method is ${\theta_{SkH}^{ZL}} = 3.26^{\circ}$.  Therefore,  the skyrmions are moving in the direction the current is applied without getting deflected much which makes it suitable for logic operation. At the magnitude of applied current = 1 x 10$^{12}$ A/m$^2$, the average velocity of the skyrmion is $\approx$ 60.24 m/s. 
  
%%%%%%%%%%%%%%%%%%%%%%%%%%%%%%%%%%%%%%%%%%%%%%%%%%%%%%%  

\subsection{Skyrmion motion in other 4dFe/Ir(111)}
Same method of nucleating skyrmions using dc magnetic field can be followed in case of all the XFe/Ir(111) material system, as reported in Ref. \cite{mukherjee2025interplay} where X = Pd, Rh, Ru, Mo.  Here, we observe skyrmion dynamics in these 4dFe/Ir(111) under the application of nano second current pulse in +z  direction.  All the material systems show similar behavior as described in case of PdFe/Ir(111).  All the skyrmions first move towards an edge of the sample depending on the direction of current applied, some of them survive and accumulate at one edge. Based on the topological charge of the sample the skyrmions show relaxed motion in clockwise or anti-clockwise direction after removing the current pulse.  One such case is explained in Fig. \ref{fig10} for RuFe/Ir(111).

\section{Conclusion and outlook}
\label{sec_conclu}

\par In conclusion, the nucleation and dynamics  of skyrmions are investigated in PdFe/Ir(111) by applying a nano second current pulse using micromagnetic simulation. Simultaneous creation of skyrmions of two opposite topological charges $\pm$1  are observed in a spin-spiral background. On the other hand, after applying a magnetic field to achieve a complete skyrmion state, current pulses are used to manipulate their arrangement, enabling logic operations such as OR and AND. We tune the direction and current density ranging from (10$^{11}$-10$^{12}$) A/m$^2$ of nano second current pulses to make different desired shapes like `T', `$\pi$', `L', `C', and `7'. We employ the concept of anisotropy barrier to regulate skyrmion dynamics in the nano structure for designing the logic gates considering two vertical chains of skyrmions.  Further we verify this method of logic gate construction in `$\pi$' shaped structure of skyrmion.

\par Beyond the PdFe/Ir(111) system, the other 4dFe/Ir(111) multilayers such as RhFe/Ir(111) exhibits similar dynamic behavior under nano second current pulses, which can subsequently be used to tailor desired configurations and implement logic-based architectures in 4d transition metal multilayers. Our detailed study provides a design protocol for reconfigurable, low-power logic architectures based on skyrmion dynamics in 4d transition metal multilayers and highlight the potential of interface-engineered anisotropy in future spintronic computation schemes.

\section*{Acknowledgement}

TM and VSNM acknowledge BITS Pilani, Hyderabad Campus for 
providing the Sharanga High-Performance Computational Facility. BS acknowledges Prime Ministers Early Career Research Grant (PMECRG) by Anusandhan National Research Foundation (ANRF) for this project with reference number ANRF/ECRG/2024/005021/PMS.

%%%%%%%%%%%%%%%%%%%%%%%%%%%%%%%%%%%%%%%%%%%%%%%%%%%%%%%
%%%%%%%%%%%%%%%%%%%%%%%%%%%%%%%%%%%%%%%%%%%%%%%%%%%%%%%
%%%%%%%%%%%%%%%%%%%%%%%%%%%%%%%%%%%%%%%%%%%%%%%%%%%%%%%

\section*{Appendix}

\begin{figure*}[ht]
\includegraphics[width=0.99\textwidth,angle=0]{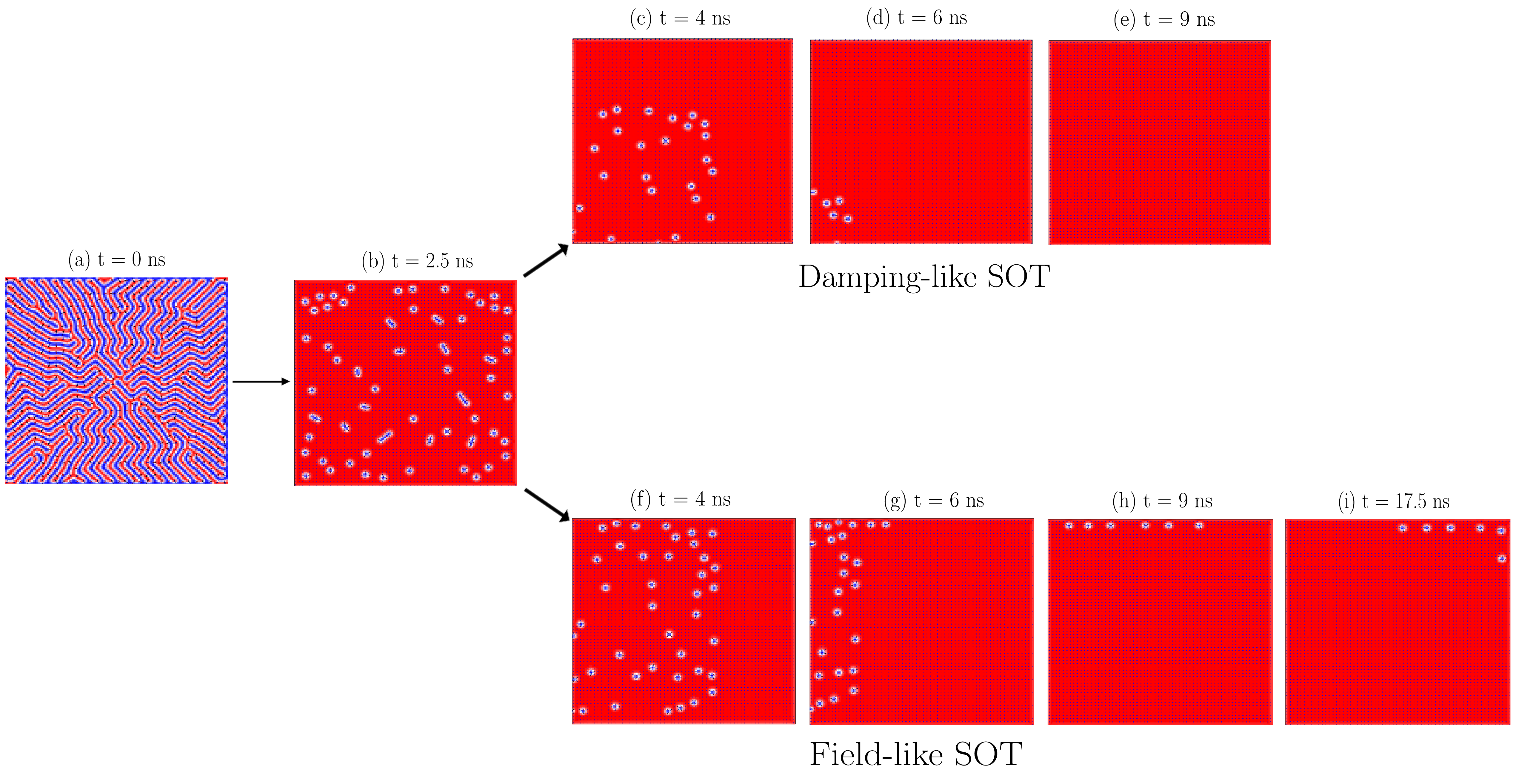}
\caption{{Current pulse of 5 $\times$ 10$^{11}$ A/m$^2$ is applied through the heavy metal layer in +x - direction for 10 ns and then switched off. The initial (a) spin-spiral ground state transforms into (b) complete skyrmion state after 2.5 ns in presence of B$_{ext}$ = 2 T, (c)-(e) analyzes the system when damping-like SOT is considered. Eventually all the skyrmions disappear within this time. Moreover, (f)-(i) demonstrates the system when only field like SOT is taken into account. (f)-(h) are snapshots of the system when the current density pulse is on, (i) skyrmions are moving in clockwise direction along the edge in the absence of any current pulse.}}
\label{sfig1}
\end{figure*}

\section{Different shapes in square geometry}

\begin{figure*}
   \includegraphics[width=0.6\textwidth,angle=0]{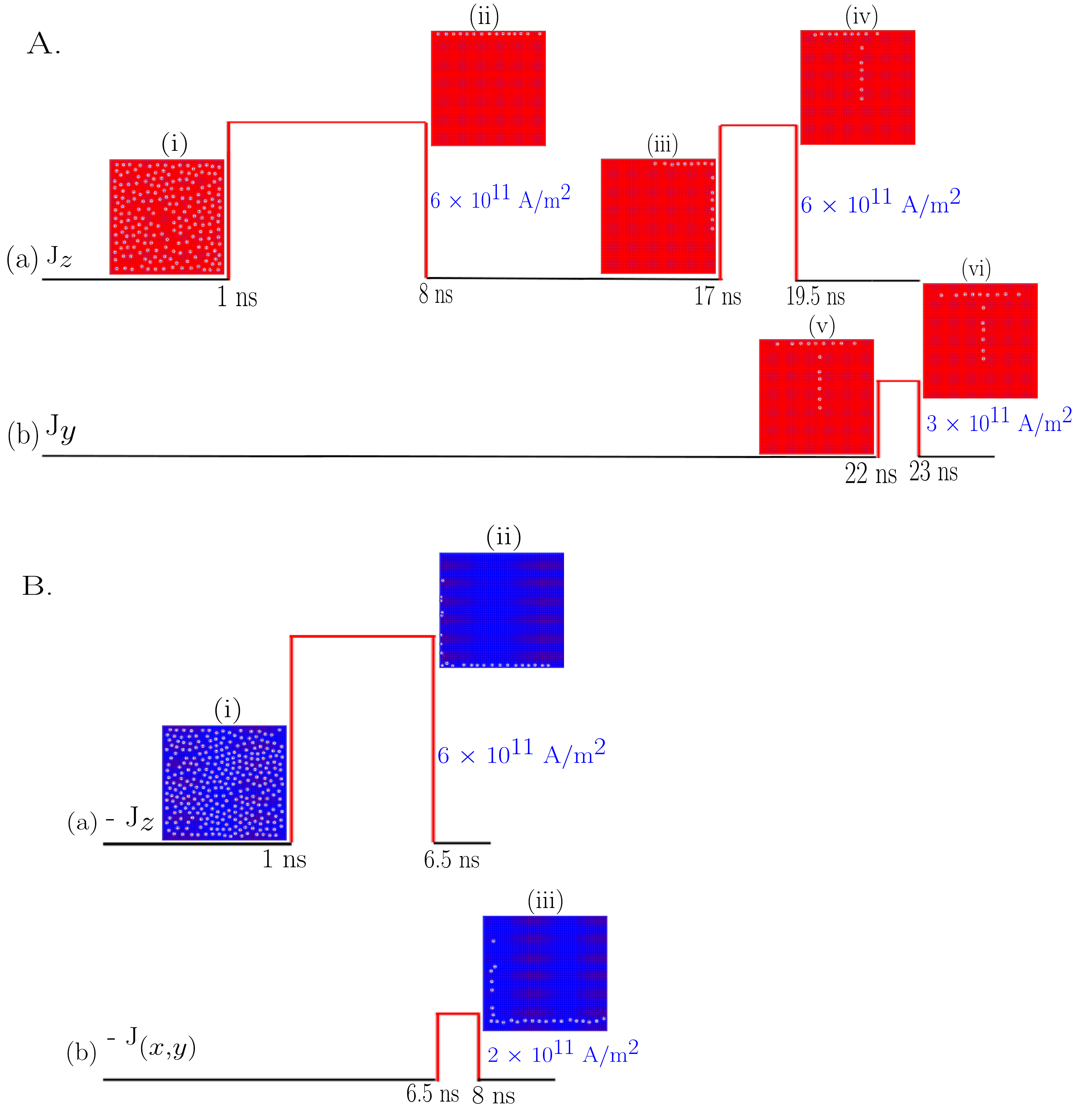}
\caption{{ Manipulating the skyrmions in the square nano-structure to form} (A) {\textbf{`T'}} Shape: (a) current pulse along +z --direction; (i) - (iv) intermediate states during `T' - Shape formation, and (b) current pulse along +y -- direction; (v) - (vi) snapshots when +y -- direction pulse is on. (B){\textbf{ `L'}} Shape: (a) - (b) current pulses are applied along -z, (-x, -y)  --directions; (i) - (iii) intermediate states during the 7 - shape formation. }
\label{sfig2}
\end{figure*}

\subsection{T shape}
\par A current pulse of 6 $\times$ 10$^{11}$ A/m$^2$ in +z direction is applied for 7 ns and then switched off for 9 ns.  A line of skyrmions gets reflected from the top-left corner and starts accumulating at the top edge while the current pulse is on, and then starts moving in a clockwise direction when the current pulse is off.  Meanwhile,  some of the skyrmions reach to the right edge and then the current pulse of 6 $\times$ 10$^{11}$ A/m$^2$ in +z direction is again switched on for 2.5 ns which moves this line of skyrmions to the left (-x direction).  At this time, the remaining top edge skyrmions show relaxed motion and come to the centre of the top edge to make the horizontal line of the `T’ shape as shown in Fig.\ref{sfig2}. At last,  3 $\times$ 10$^{11}$ A/m$^2$ is applied in +y direction for 1 ns to move the  `T'  at the centre of the square nano-structure.

\subsection{L shape}
\par Now, the skyrmions of +z core magnetization are taken into consideration.  Current pulse of 6 $\times$ 10$^{11}$ A/m$^{2}$ is applied in the -z direction for 5.5 ns and some skyrmions reverse back from down-left corner and start accumulating along the lower edge. Two lines, one at the left and another at the bottom are obtained. Then 2 $\times$ 10$^{11}$ A/m$^{2}$ of current density pulse is applied in (-x, -y) direction for 1.5 ns to move the whole `L' structure near the centre of the sample as shown in Fig.\ref{sfig2}.

\section{Effect of host geometry on skyrmion manipulation other than square}

\par
\begin{figure*}
\includegraphics[width=0.85\textwidth,angle=0]{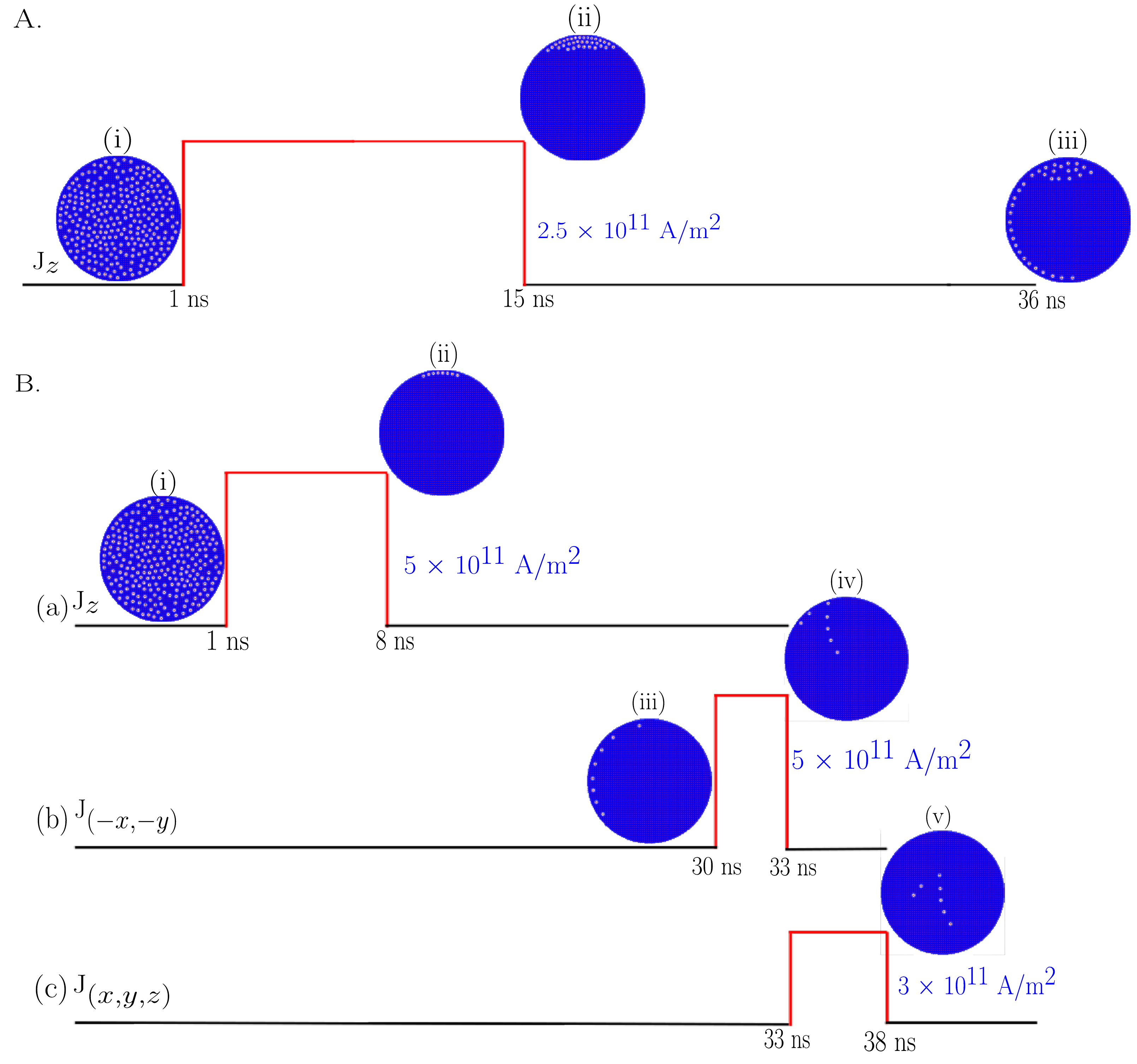}
\caption{{Manipulating the skyrmions in the circular nano-structure to form} (A) {\textbf{`C'}} Shape: (a) current pulse along +z --direction; (i) - (iii) intermediate states during C - Shape formation.  (B){\textbf{ `7'}} Shape: (a) - (c) current pulses are applied along +z, (-x, -y) and (x, y, z) --directions; (i) - (v) intermediate states during the 7 - shape formation.}
\label{sfig3}
\end{figure*}

 Skyrmion nucleation in PdFe/Ir(111) under a dc magnetic field ($\vec{B}_{ext}$) is investigated in two geometries — a square of (200 $\times$ 200) nm$^2$ and a circle with a diameter of 200 nm to assess the influence of sample shape.  The number of skyrmions at a fixed value of an external magnetic field varies with the geometric shape as the physical space available also changes with the geometry. We observe the formation of 58 skyrmions in the square nano-structure and 42 skyrmions in the circular nano structure under an external magnetic field $\vec{B}_{ext}=(0, 0, 2)$ T. However, the system transforms from a mixed spin-spiral + skyrmion state to complete skyrmion state at the same value of $\vec{B}$$_{ext}$ = (0, 0, 1.7) T. With the increasing value of $\vec{B}$$_{ext}$, the number density shows rapid increment and finally the system changes to a single domain ferromagnet at 4.5 T.  Figure \ref{sfig3} presents a comparative analysis of nucleation of skyrmions in circular nano-structures respectively at $\vec{B}_{ext}= (0, 0, 3)$ T.

\par `C' shape formation is observed in circular nano strcuture.  An external dc magnetic field ranging from 1.7 T to 4.5 T can nucleate a significant number of skyrmions in circular disc type structure of 200 nm diameter and 1 nm thickness.  We observe, the skyrmions follow similar dynamics like in the square nano structure.  A stable phase of total 250 isolated skyrmions of core magnetization along +z direction is obtained after applying 3 T of magnetic field in -z direction.  A current pulse of 2.5 $\times$ 10$^{11}$ A/m$^2$ is applied in the +z direction for 14 ns.  All the skyrmions start moving to +x direction. Some of them get reflected back from the edge and start accumulating at the top arc of the nano strcture.  Then the current is switched off and the skyrmions start moving along the edge and we keep the simulation on for another 21 ns and the final output is a `C'.

\par Now, current pulse is increased to 5 $\times$ 10$^{11}$ A/m$^2$ and applied in +z direction for 7 ns.  As the magnitude of the current is higher than earlier, a lesser number, a total of 8 skyrmions survive this time. Then, the current is switched off for 22 ns.  Meanwhile, the skyrmions start moving along the edge. Then a current pulse of 5 $\times$ 10$^{11}$ A/m$^2$ is applied in the (-x, -y) direction for 3 ns.  During this time, the skyrmions move in (+x, +y) direction. One skyrmion got annihilated at the edge and others form a `7' due to the interaction between themselves and with the sample edge.  To move the structure `7' at the centre of the sample, we further apply a current density pulse of 3 $\times$ 10$^{11}$ A/m$^2$ in (+x, +y, +z) direction for 5 ns.

\section{Logic Gate architectures with `$\pi$' -- shape}

\par
\begin{figure*}
 \includegraphics[width=0.99\textwidth,angle=0]{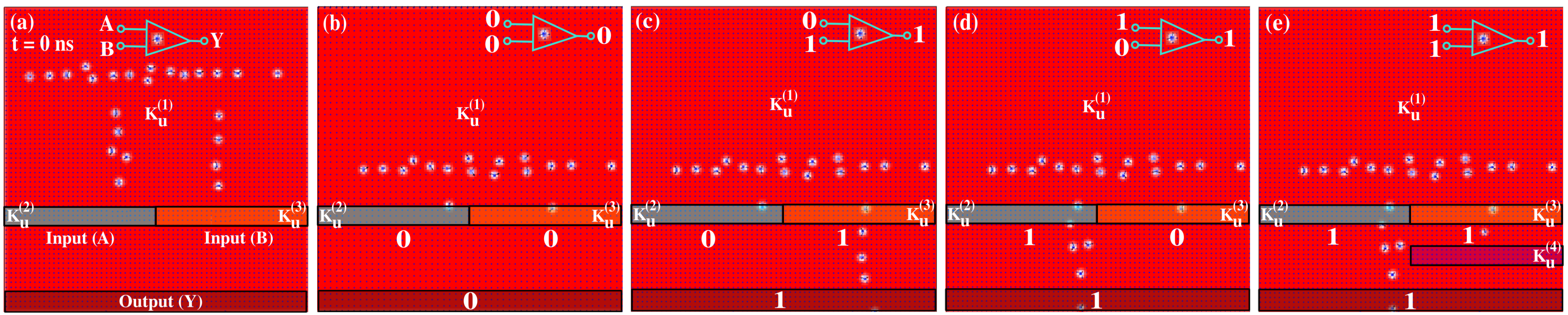}
\caption{OR logic gate functionality using $\pi$'.  Anisotropy constants K$_u^{(2)}$ = 1.5K$_u^{(1)}$ or, 2K$_u^{(1)}$ (placed at (-50 nm, -40 nm)); K$_u^{(3)}$ = 1.5K$_u^{(1)}$ or, 2K$_u^{(1)}$ (placed at (50 nm, -40 nm)), and K$_u^{(4)}$ = 2K$_u^{(1)}$ (placed at (50 nm or -50 nm, -50 nm)) are assumed. Snapshots of the logic operations (a) at t = 0 ns, (b) 0 + 0 = 0, (c) 0 + 1 = 1, (d) 1 + 0 = 0, and (e) 1 + 1 = 1.  } 
\label{sfig4}
\end{figure*}

\begin{figure*}
\includegraphics[width=0.99\textwidth,angle=0]{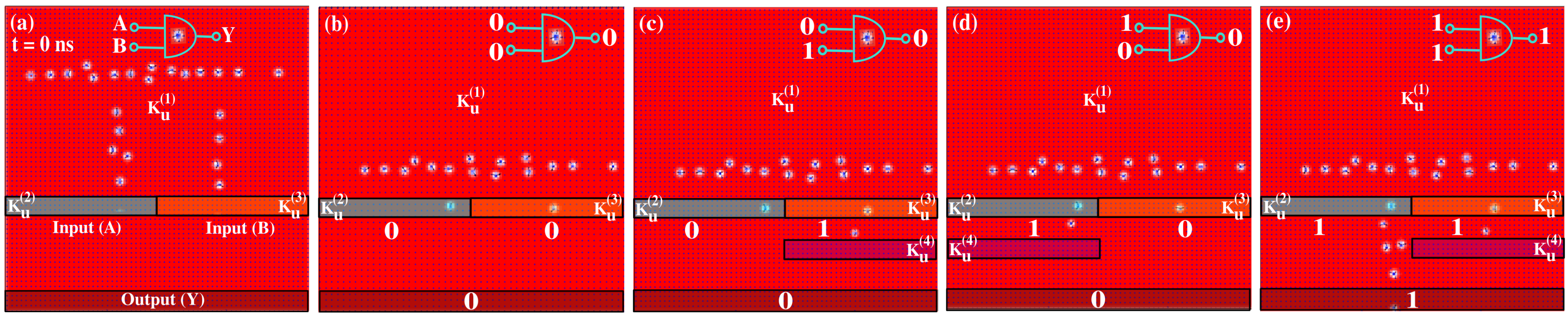}
\caption{AND logic gate functionality using $\pi$'. Three anisotropy barriers of anisotropy constants K$_u^{(2)}$ = 1.5K$_u^{(1)}$ or, 2K$_u^{(1)}$; K$_u^{(3)}$ = 1.5K$_u^{(1)}$ or, 2K$_u^{(1)}$ and K$_u^{(4)}$ = 2K$_u^{(1)}$ are considered. To implement the logic operation of an AND gate, these barriers are assumed to occupy positions similar to those used in the OR gate configuration. Snapshots of the logical multiplications  (a) at t = 0 ns, (b) 0 . 0 = 0, (c) 0 . 1 = 0, (d) 1 . 0 = 0, and (e) 1 . 1 = 1.   } 
\label{sfig5}
\end{figure*}

 A `$\pi$' shape is formed in the square nano structure after applying 3 T external magnetic field in the +z  direction.   Proper adjustment the magnitude and direction of the nano second current density pulse are done to make logic gates OR and AND as depicted in Fig. \ref{sfig4} and \ref{sfig5} respectively.  Three anisotropy barriers of length = 100 nm, width = 10 nm are considered to make the logic gate working.  Anisotropy constants K$_u^{(2)}$ = 1.5K$_u^{(1)}$ or, 2K$_u^{(1)}$ (placed at (-50 nm, -40 nm)); K$_u^{(3)}$ = 1.5K$_u^{(1)}$ or, 2K$_u^{(1)}$ (placed at (50 nm, -40 nm)), and K$_u^{(4)}$ = 2K$_u^{(1)}$ (placed at (50 nm or -50 nm, -50 nm)) are assumed. A current pulse of 1 $\times$ 10$^{12}$ A/m$^2$ is applied in the +y direction to move the $\pi$ structure.

\bibliography{pdfeir}{}

\clearpage

\end{document}